\documentclass[floatfix,aps,prl,showpacs,amsmath,nofootinbib,preprintnumbers,twocolumn]{revtex4}

\usepackage{graphicx}
\usepackage{bm}

\def\half{{1\over 2}}

\def\half{{1\over 2}}
\def\({\left(}
\def\){\right)}
\def\[{\left[}
\def\]{\right]}

\def\e{\begin{equation}}
\def\q{\end{equation}}
\def\m{\begin{eqnarray}}
\def\n{\end{eqnarray}}

\begin{document}

\title{Cosmological constant, inflation and no-cloning theorem}


\author{Qing-Guo Huang$^{a}$\footnote{huangqg@itp.ac.cn} and Feng-Li Lin$^{b,c}$\footnote{linfengli@phy.ntnu.edu.tw}\footnote{On leave from National Taiwan Normal University.}}
\affiliation{${}^{a}$ State Key Laboratory of Theoretical Physics, Institute of Theoretical Physics, Chinese Academy of Science, Beijing 100190, People's Republic of China,}\affiliation{${}^{b}$ Department of Physics, Massachusetts Institute of Technology, Cambridge, Massachusetts 02139, USA}
\affiliation{${}^{c}$ Department of Physics, National Taiwan Normal University, Taipei, 116, Taiwan}


\begin{abstract}

From the viewpoint of no-cloning theorem we postulate a relation between the current accelerated expansion of our universe and the inflationary expansion in the very early universe. It implies that the fate of our universe should be in a state with accelerated expansion. Quantitatively we find that the no-cloning theorem leads to a lower bound on the cosmological constant which is compatible with observations.

\end{abstract}


\maketitle


Cosmic acceleration \cite{Riess:1998cb,Perlmutter:1998np} is one of the most important discoveries in the past decades. It strongly suggests that our universe has a small positive cosmological constant 
\e
\Lambda\simeq 1.18\times  10^{-123} M_{pl}^4,
\q 
where $M_{pl}=1/\sqrt{G}=1.22\times 10^{19}$ GeV is the Planck mass. If the universe is described by an effective local quantum field theory up to the Planck scale, we would expect a cosmological constant of order of $M_{pl}^4$ which is larger than the observed value by roughly 120 orders of magnitude \cite{Weinberg:1988cp}. How to explain such a small but non-zero positive value is an outstanding theoretical challenge. It is the so called cosmological constant problem.

Actually the cosmic acceleration does not only happen in the late time universe, but also occurs in the very early universe. Before hot Big Bang our universe was proposed to be in an inflationary phase \cite{Guth:1980zm} which is a period of nearly exponential growth in the very early universe. A spatially flat universe is strongly supported by cosmological observations \cite{Komatsu:2010fb}. Roughly speaking, inflation should last not less than 60 e-foldings (or equivalently, our universe expands not less than $10^{26}$ times during the inflationary period) in order to naturally explain the flatness of the universe. The geometry of inflationary universe can be taken as a quasi-de Sitter(dS) space which has an event horizon like that for a back hole. An observer in dS space sees the surrounding spacetime as a finite closed cavity bounded by a horizon with size denoted by $R_I$, and the cavity is described by a thermal ensemble at temperature $T=1/2\pi R_I$ \cite{Gibbons:1977mu}. This thermal ensemble has a finite entropy given by 
\m
S_I={\pi R_I^2 \over \ell_p^2},
\n
where $\ell_p=\sqrt{G}$ is the Planck length. 

As for any closed system information can never be lost but can be scrambled and thermalized at the hot horizon. Similar to the black hole complementarity principle \cite{Susskind:1993if}, for dS space the complementarity principle \cite{Banks:2001yp} says that to an observer who never crosses the horizon, the horizon can absorb, thermalize and re-emit all information that falls on it. It can be also interpreted as that there is no loss of information. 

Now we consider that an observer, call him Bob, stays at a point of $r=0$ and his partner, called Alice stays at a point with comoving coordinate $r_c$ away from Bob. The physical distance between Alice and Bob is given by $r_p=a(t)r_c$ where $a(t)$ is the scale factor. The dynamics of inflationary universe is governed by an effective cosmological constant $\Lambda_I$, and the scale factor exponentially grows up: $a(t)\sim e^{H_I t}$, where the Hubble parameter $H_I$ is related to $\Lambda_I$ by 
\m
H_I=\sqrt{8\pi \Lambda_I\over 3 M_{pl}^2}. 
\n
The horizon size of such an inflationary universe is nothing but $R_I=H_I^{-1}$.
At the beginning, Alice is assumed to stay inside Bob's event horizon. She then carries an encoded qubit with her and crosses Bob's event horizon at a moment denoted by $t_c$ due to the accelerating expansion of the universe. Based on the complementarity principle for dS space, the information of such a qubit is absorbed in the event horizon and Bob can reconstruct it  after the moment of $t_c+t_*$ by collecting the evaporations from the event horizon. If $t_*$ is not infinitely long, there would be a possible conflict with no-cloning theorem of quantum mechanics as we will describe below. In \cite{Danielsson:2002td}, the authors argued that the minimum time needed to measure the information in the radiation in dS space is order of $R_I^3/\ell_p^2$. Recently Susskind in \cite{Susskind:2011ap} conjectured that dS space is a fast scrambler \cite{ft1} whose scrambling time is 
\e
t_*=\alpha R_I \ln {R_I\over \ell_p}, 
\label{sct}
\q
which is much shorter than $R_I^3/\ell_p^2$, where $\alpha$ is a constant of order unity.

In a universe with everlasting inflation Bob can never commute with Alice any more after she crosses the event horizon. Thus Bob has no chance to receive such a qubit sent by Alice even in the form of a photon whenever Alice sends the qubit to Bob after she crosses the horizon. In this case Bob cannot clone a qubit. The case of eternal inflation in landscape was discussed in \cite{Nomura:2011dt}. However, in our universe  inflation must end at a moment ($t_{end}$). Otherwise, there would be no matter and radiation we observed today. In this case Bob can reconstruct the qubit carried by Alice from the Gibbon-Hawking radiations during the inflationary era if the inflation ends after $t_c+t_*$. The danger is that Bob has a chance to receive the qubit sent from Alice sooner or later in a matter and/or radiation dominated universe. It implies that Bob can clone a qubit in such a universe. 

On the other hand, both quantum superposition and unitarity are the fundamental principles of quantum mechanics. They forbid the creation of identical copies of an arbitrary unknown quantum state. It was stated as the no-cloning theorem \cite{wootters:1982}. Once one can clone a qubit, the unitarity in the quantum mechanics will be violated.  In order to solve the puzzle of cloning a qubit in an inflationary universe followed by matter and/or radiation dominated era, we postulate that there should be a cosmological constant which will govern the dynamics of the late time universe.

In following part of this letter, we will figure out a quantitative relation between the inflation scale and the value of cosmological constant driving the late time cosmic acceleration in detail. We need to stress that the main point of this letter is not to give an answer to the cosmological constant problem. The problem we concern is why there should be a cosmological constant which dominates the late time universe.

Let's recall the previous ``Alice and Bob" thought experiment in an inflationary universe. Because dS space is a fast scrambler, the minimal e-folding number for Bob to reconstruct the qubit carried by Alice from the Gibbons-Hawking radiations during inflation is 
\e
N_q=H_I t_*=\alpha \ln {R_I\over \ell_p}\sim \ln S_I
\label{nq} 
\q
which is expected to be ${\cal O}(10)$. As we know, the minimal number of e-foldings for solving the flatness problem is around 60 and the new inflation usually lasts much longer than hundreds e-foldings \cite{newinflation}. Thus usually one can expect that Bob has enough time to reconstruct the qubit carried by Alice during the inflationary period.

At the end of inflation, the proper distance between Alice and Bob is given by 
\e\label{Lq0}
L_q=e^{H_I(t_{end}-t_c)} H_I^{-1}.
\q
Therefore the most dangerous case is that Alice crosses the event horizon at the momentum of $t_c=t_{end}-t_*$ and then
\e
L_q = \({M_{pl}\over H_I}\)^\alpha H_I^{-1}\sim S_I^{\alpha/2} H_I^{-1}.
\label{lq}
\q
Now whether Bob can clone such a qubit is translated into whether Bob can receive it from Alice at a moment of $t_f$ ($t_f$ is finite) after the end of inflation.

After the end of inflation, the vacuum energy governing the dynamics of inflation decays into radiations and matters. The radiation energy density goes like $a^{-4}$ and the matter energy density goes like $a^{-3}$. The evolution of scale factor in a radiation or matter dominated universe is $a(t)\sim t^p$, where $p=1/2$ and $p=2/3$ correspond to radiation dominated and matter dominated era respectively. For simplicity, we normalize the scale factor at the end of inflation to be unity, namely $a(t_{end})=1$, and then $a(t)=(t/t_{end})^p$. The comoving  distance between Bob and Alice's qubit at the end of inflation is still given by that in Eq.~(\ref{lq}). If our universe is always dominated by radiation and/or matter, the comoving distance travelled by Alice's photon qubit from $t_{end}$ to $t_f$ is 
\m
L_{r,m}=\int_{t_{end}}^{t_f} {dt\over a(t)} \sim t_{end} \({t_f\over t_{end}}\)^{1-p}. 
\n
Here we consider $p<1$ which is valid for matter and/or radiation dominated universe. For $t_f\rightarrow \infty$, $L_{r,m}\rightarrow \infty$ which implies that the qubit can travel to any place in the whole space. More precisely, one can easily find that $L_{r,m}>L_q$ if
\m
t_f>t_{end} \[{1\over H_I t_{end}} \({M_{pl}\over H_I}\)^\alpha\]^{1\over 1-p}.
\n
It indicates that Bob can clone a qubit if the inflation lasts longer than the scrambling time in an inflationary universe, followed by matter and/or radiation dominated era.

In order to avoid the above violation of no-cloning theorem, we suggest that there should be a positive cosmological constant $\Lambda$ which governs the late-time evolution of our universe. Intuitively, such a cosmological constant implies that we live in a space-time that will asymptotically tend to dS space. The horizon size of this asymptotical dS space is given by   
\e
H_{\Lambda}^{-1}=\sqrt{3M_{pl}^2 \over 8\pi \Lambda}.
\q
Because Alice's qubit was inflated further away from Bob by the late time cosmic acceleration, Bob cannot receive the qubit sent from Alice. The no-cloning theorem will then be preserved as long as the cosmological constant is large enough so that the distance between Alice and Bob at the end of inflation is not less than the event horizon size of the asymptotical dS space, namely $L_q\gtrsim H_{\Lambda}^{-1}$. From this inequality, we get 
\m
\Lambda \gtrsim S_I^{-\alpha} \Lambda_I. 
\label{est}
\n
Because $H_I/M_{pl}\lesssim 10^{-5}$ ($S_I>10^{10}$) \cite{Komatsu:2010fb} and $\alpha\sim {\cal O} (1)$, the smallness of cosmological constant dominating the later time universe compared to the effective vacuum energy driving inflation can be understood by the the huge dS entropy associated with the inflation if the inequality in (\ref{est}) is saturated.

The above results bases on a quite rough estimation. It can illustrate the main point of physics. From now on we will try to make a more accurate estimation. After inflation ends, our universe is filled with relativistic particles which are taken as radiation. With the cosmic expansion, the radiation energy density decreases very fast and the non-relativistic matter starts to become dominant at the time of $t_{eq}$. Because the energy density of cosmological constant is a constant, it will be dominant and drive the late time cosmic acceleration sooner or later. The time when the transition from matter dominated era to accelerating expansion takes place is denoted by $t_{ac}$. Therefore the maximum comoving  distance travelled by the photon qubit sent from Alice takes the form 
\m
L_h&\simeq& \int_{t_{end}}^{t_{eq}} {dt\over (t/{t_{end}})^{1/2}}+\int_{t_{eq}}^{t_{ac}} {dt\over (t_{eq}/{t_{end}})^{1/2}(t/{t_{eq}})^{2/3}}\nonumber \\
&+&\int_{t_{ac}}^{\infty} {dt\over (t_{eq}/{t_{end}})^{1/2}(t_{ac}/{t_{eq}})^{2/3}e^{H_\Lambda (t-t_{ac})}}\nonumber \\
&\simeq& \# {T_{eq}\over T_{end}} H_\Lambda^{-1},
\label{lh}
\n
where 
\e
\#=c_1 {M_{pl}H_\Lambda\over 5 T_{eq}^2}+c_2 z_{eq}^{-1},
\q
$c_1$ and $c_2$ are the constants of order unity, $T_{end}$ and $T_{eq}$ are the temperatures of radiation at the time of $t_{end}$ and $t_{eq}$ respectively, and $z_{eq}$ is the redshift corresponding to $t_{eq}$. A factor of $\# \ {T_{eq}/ T_{end}} $ is missed in the former intuitive estimation. In the limit of $\Lambda\rightarrow 0$, $\#\simeq  c_2 z_{eq}^{-1}$. Assuming that the vacuum energy during inflation instantaneously decays into radiation, we roughly have $T_{end}\simeq \sqrt{M_{pl}H_I}$. No-cloning theorem requires $L_q\gtrsim L_h$ which yields 
\m
{\Lambda\over M_{pl}^4}\gtrsim {\Lambda_c\over M_{pl}^4}\equiv \#^2 \({T_{eq}\over M_{pl}}\)^2 \({\Lambda_I\over M_{pl}^4}\)^{\alpha+\half}. 
\label{blambda}
\n
From no-cloning theorem, we find that the cosmological constant driving the late time cosmic acceleration is bounded from below by $\Lambda_c$ which is determined by the inflation energy scale.

The cosmological observations \cite{Komatsu:2010fb} indicate that $T_{eq}\simeq 6.2\times 10^{-29} M_{pl}$,  $H_\Lambda\simeq 9.94\times 10^{-62} M_{pl}$, $z_{eq}\simeq 3196$ in our universe, and then we get $\#\sim 10^{-3}$. Taking the value of cosmological constant as an input, we obtain an upper bound on the inflation scale, namely
\m
{\Lambda_I^{1/4}\over M_{pl}}\lesssim (3\times 10^{-61})^{1\over 4\alpha+2}. 
\n
If $\alpha=1$, $\Lambda_I^{1/4}\lesssim 10^9$ GeV and a Grand Unification Theory (GUT) scale $(10^{16}\ \rm{GeV})$ inflation does not survive. If $\alpha\geq 5$, GUT scale inflation still survives.


Summary and Discussion. First of all, we need to stress that we do not solve the cosmological constant problem in this letter. We only postulate a lower bound on the late time cosmological constant for protecting the unitarity of quantum theory if inflation in the early universe lasts longer than the scrambling time. However, an upper bound on it is still absent \cite{Weinberg:1987dv}. Ones believe that the quantum theory of gravity is needed before solving the cosmological constant problem. Unfortunately, the quantum gravity theory has not been well-established. In this letter, we get some new insights into the cosmological constant by taking into account the quantum effects of dS space which may encode some important properties of quantum gravity, such as holography and complementarity.  

Whether there is information loss in a strong gravitational system, such as black hole, is a long-standing puzzle. Nowadays many people believe that there is no information loss and the unitarity is still preserved once the full quantum theory of gravity is considered. A similar ``Alice and Bob thought experiment" for a black hole was discussed in \cite{Susskind:1993mu,Hayden:2007cs,Sekino:2008he} where they found that Bob can not catch up the qubit before he hits the singularity inside the black hole. However there is not a singularity in the late-time universe. Here we propose that the event horizon due to a positive cosmological constant can protect the no-cloning theorem. 

Actually a similar discussion is also applicable for the more general dark energy model and what we need to do is just to replace $H_{\Lambda}$ in Eq.~(\ref{lh}) by the Hubble parameter at the time of transition from decelerated expansion to accelerated expansion. But the fate of universe should be in a state with accelerated expansion. Otherwise, Bob can clone the qubit sooner or later and the unitarity will be violated.

Another possibility is that the total number of e-folds is bounded from above by $N_q$ in Eq.~(\ref{nq}) and then we do not need to worry about the violation of no-cloning theorem at all. But it is quite difficult to construct a realistic inflation model with a small total number of e-folds \cite{Linde:2003hc}. If it is the case, the spatial curvature of our universe might be detected in the future.



\vspace{5mm}

\noindent {\bf Acknowledgments}

QGH would like to thank J.~X.~Lu, C.~P.~Sun, P.~J.~Yi and S.~Yi for useful conversations.  FLL thanks Wei Li for helpful discussions, and the support of NCTS. QGH is supported by the project of Knowledge Innovation Program of Chinese Academy of Science and a grant from NSFC (grant NO. 10975167). FLL is supported by Taiwan's NSC grants (grant NO. 100-2811-M-003-011 and 100-2918-I-003-008).




\begin{thebibliography}{99}
\frenchspacing



\bibitem{Riess:1998cb}
  A.~G.~Riess {\it et al.}  [Supernova Search Team Collaboration],
  Astron.\ J.\  {\bf 116}, 1009 (1998)
  [arXiv:astro-ph/9805201].
  
\bibitem{Perlmutter:1998np}
  S.~Perlmutter {\it et al.}  [Supernova Cosmology Project Collaboration],
  Astrophys.\ J.\  {\bf 517}, 565 (1999)
  [arXiv:astro-ph/9812133].    


\bibitem{Weinberg:1988cp}
  S.~Weinberg,
  Rev.\ Mod.\ Phys.\  {\bf 61}, 1-23 (1989).

\bibitem{Guth:1980zm}
  A.~H.~Guth,
  Phys.\ Rev.\  D {\bf 23}, 347 (1981).


\bibitem{Komatsu:2010fb}
  E.~Komatsu {\it et al.}  [WMAP Collaboration],
  Astrophys.\ J.\ Suppl.\  {\bf 192}, 18 (2011)
  [arXiv:1001.4538 [astro-ph.CO]].
  
  


\bibitem{Gibbons:1977mu}
  G.~W.~Gibbons and S.~W.~Hawking,
  Phys.\ Rev.\  D {\bf 15}, 2738 (1977).



\bibitem{Susskind:1993if}
  L.~Susskind, L.~Thorlacius and J.~Uglum,
  Phys.\ Rev.\  D {\bf 48}, 3743 (1993)
  [arXiv:hep-th/9306069].



\bibitem{Banks:2001yp}
  T.~Banks and W.~Fischler,
  arXiv:hep-th/0102077; 
  T.~Banks, W.~Fischler and S.~Paban,
  JHEP {\bf 0212}, 062 (2002)
  [arXiv:hep-th/0210160]; 
  L.~Dyson, J.~Lindesay and L.~Susskind,
  JHEP {\bf 0208}, 045 (2002)
  [arXiv:hep-th/0202163]; 
  L.~Susskind,
  arXiv:hep-th/0204027; 
  L.~Dyson, M.~Kleban and L.~Susskind,
  JHEP {\bf 0210}, 011 (2002)
  [arXiv:hep-th/0208013]; 
  M.~K.~Parikh, I.~Savonije and E.~P.~Verlinde,
  Phys.\ Rev.\  D {\bf 67}, 064005 (2003)
  [arXiv:hep-th/0209120].
  
  
\bibitem{Danielsson:2002td} 
  U.~H.~Danielsson, D.~Domert and M.~E.~Olsson,
  Phys.\ Rev.\ D {\bf 68}, 083508 (2003)
  [hep-th/0210198]; 
  U.~H.~Danielsson and M.~E.~Olsson,
  JHEP {\bf 0403}, 036 (2004)
  [hep-th/0309163].


\bibitem{Susskind:2011ap}
  L.~Susskind,
  arXiv:1101.6048 [hep-th].  

\bibitem{ft1}
For a finite size system whose total number of degrees of freedom scales with a parameter $N$, such a system is called ``fast scrambler" if the time for the thermalization saturates $t_*\gtrsim T^{-1}\ln N$. Here $T$ is the temperature.


\bibitem{Nomura:2011dt} 
  Y.~Nomura,
  JHEP {\bf 1111}, 063 (2011)
  [arXiv:1104.2324 [hep-th]].
  

\bibitem{wootters:1982}
W.~K.~Wootters and W.~H.~Zurek, 
 Nature 299 (1982), pp. 802.


\bibitem{newinflation} 
  A.~D.~Linde,
  Phys.\ Lett.\ B\ {\bf 108}, 389  (1982);
   A.~Albrecht and P.~J.~Steinhardt,
  Phys.\ Rev.\ Lett.\ \ {\bf 48}, 1220  (1982).



\bibitem{Susskind:1993mu}
  L.~Susskind and L.~Thorlacius,
  Phys.\ Rev.\  D {\bf 49}, 966 (1994)
  [arXiv:hep-th/9308100].
  
\bibitem{Hayden:2007cs}
  P.~Hayden and J.~Preskill,
  JHEP {\bf 0709}, 120 (2007)
  [arXiv:0708.4025 [hep-th]].


\bibitem{Sekino:2008he}
  Y.~Sekino and L.~Susskind,
  JHEP {\bf 0810}, 065 (2008)
  [arXiv:0808.2096 [hep-th]].  
  

\bibitem{Weinberg:1987dv}
A possible way to get an upper bound on the cosmological constant is to consider the condition for the formation of galaxies. For example, see
  S.~Weinberg,
  Phys.\ Rev.\ Lett.\  {\bf 59}, 2607 (1987).
  
  
\bibitem{Linde:2003hc} 
  A.~D.~Linde,
  JCAP {\bf 0305}, 002 (2003)
  [astro-ph/0303245].  
  
  
  
  
\end{thebibliography}
\end{document}